\documentclass [prb,superscriptaddress, showpacs, twocolumn]{revtex4-1}
\usepackage{graphicx}
\usepackage{dcolumn}
\usepackage{bm}

\makeatletter

\newcommand{\Rmnum}[1]{\expandafter\@slowromancap\romannumeral #1@}
\makeatother

\begin{document}

\title{Magnetic structure of an incommensurate phase of La-doped BiFe$_{0.5}$Sc$_{0.5}$O$_3$: Role of antisymmetric exchange interactions}

\author{D. D. Khalyavin}
\email{email: dmitry.khalyavin@stfc.ac.uk}
\affiliation{ISIS facility, STFC, Rutherford Appleton Laboratory, Chilton, Didcot, Oxfordshire, OX11-0QX,UK}
\author{A. N. Salak}
\email{email: salak@ua.pt}
\affiliation{Department of Materials and Ceramic Engineering/CICECO, University of Aveiro, 3810-193 Aveiro, Portugal}
\author{A. B. Lopes}
\affiliation{Department of Materials and Ceramic Engineering/CICECO, University of Aveiro, 3810-193 Aveiro, Portugal}
\author{N. M. Olekhnovich}
\affiliation{Scientific-Practical Materials Research Centre of NAS of Belarus, Minsk, 220072, Belarus}
\author{A. V. Pushkarev}
\affiliation{Scientific-Practical Materials Research Centre of NAS of Belarus, Minsk, 220072, Belarus}
\author{Yu. V. Radyush}
\affiliation{Scientific-Practical Materials Research Centre of NAS of Belarus, Minsk, 220072, Belarus}
\author{E. L. Fertman}
\affiliation{B. Verkin Institute for Low Temperature Physics and Engineering of NAS of Ukraine, 
Kharkov, 61103, Ukraine}
\author{V. A. Desnenko}
\affiliation{B. Verkin Institute for Low Temperature Physics and Engineering of NAS of Ukraine, 
Kharkov, 61103, Ukraine}
\author{A. V. Fedorchenko}
\affiliation{B. Verkin Institute for Low Temperature Physics and Engineering of NAS of Ukraine, 
Kharkov, 61103, Ukraine}
\author{P. Manuel}
\affiliation{ISIS facility, STFC, Rutherford Appleton Laboratory, Chilton, Didcot, Oxfordshire, OX11-0QX,UK}
\author{A. Feher}
\affiliation{Faculty of Science, P. J. \v{S}af\'{a}rik University in Ko\v{s}ice, Ko\v{s}ice, 04154, Slovakia }
\author{J. M. Vieira}
\affiliation{Department of Materials and Ceramic Engineering/CICECO, University of Aveiro, 3810-193 Aveiro, Portugal}
\author{M. G. S. Ferreira}
\affiliation{Department of Materials and Ceramic Engineering/CICECO, University of Aveiro, 3810-193 Aveiro, Portugal}
\date{\today}

\begin{abstract}
A 20 \% substitution of Bi with La in the perovskite Bi$_{1-x}$La$_x$Fe$_{0.5}$Sc$_{0.5}$O$_3$ system obtained under high-pressure and high-temperature conditions has been found to induce an incommensurately modulated structural phase. The room temperature X-ray and neutron powder diffraction patterns of this phase were successfully refined using the $Imma(0,0,\gamma )s00$ superspace group ($\gamma =0.534(3)$) with the modulation applied to Bi/La- and oxygen displacements. The modulated structure is closely related to the prototype antiferroelectric structure of PbZrO$_3$ which can be considered as the lock-in variant of the latter with $\gamma =0.5$. Below $T_\textrm{N} \sim 220$ K, the neutron diffraction data provide evidence for a long-range $G$-type antiferromagnetic ordering commensurate with the average $Imma$ structure. Based on a general symmetry consideration, we show that the direction of the spins is controlled by the antisymmetric exchange imposed by the two primary structural distortions, namely oxygen octahedral tilting and incommensurate atomic displacements. The tilting is responsible for the onset of a weak ferromagnetism, observed in magnetization measurements, whereas the incommensurate displacive mode is dictated by the symmetry to couple a spin-density wave. The obtained results demonstrate that antisymmetric exchange is the dominant anisotropic interaction in Fe$^{3+}$-based distorted perovskites with a nearly quenched orbital degree of freedom. 
\end{abstract}


\maketitle

\section{Introduction}

\indent Perovskite materials derived from the well-known multiferroic BiFeO$_3$ by various substitutions exhibit a variety of structural phases with interesting properties and improved functionality.\cite{ref:1,ref:2,ref:3,ref:4,ref:5,ref:6,ref:7,ref:8,ref:9,ref:10,ref:11,ref:12} The electronic degree of freedom related to the lone pair nature of Bi$^{3+}$ usually results in polar/antipolar atomic displacements in these materials. The displacements are often coupled to oxygen octahedral tilting. Both types of distortions define the symmetry of the perovskite lattice and control electric and magnetic properties as well as a cross-coupling between them. The energy landscape of some Bi-containing compositions consists of several almost degenerate phase states that can be switched by relatively small perturbations.\cite{ref:12,ref:13,ref:14,ref:15,ref:16} This offers a unique opportunity to study the structure-properties relationship using distinct structural modifications of the same material. It has been recently shown that the metastable perovskite BiFe$_{0.5}$Sc$_{0.5}$O$_3$ can be stabilized in two different polymorphs via an irreversible behavior under heating/cooling thermal cycling.\cite{ref:17} As-prepared BiFe$_{0.5}$Sc$_{0.5}$O$_3$ ceramics obtained by quenching under high pressure were characterized by a complex antipolar structure with the $Pnma$ symmetry and the $\sqrt{2}a_p \times 4a_p \times 2 \sqrt{2} a_p$ type superstructure ($a_p$ is the pseudocubic unit cell). Hereafter we use $p$-subscript to denote the pseudocubic setting. The ground state atomic configuration, however, could not be deduced unambiguously based on the experimental data due to existence of two non-equivalent $Pnma$ isotropy subgroups different by the origin choice and indistinguishable in the refinement procedure. This configuration was determined theoretically using state of the art density functional algorithms for structure relaxation.\cite{ref:18} On heating, the antipolar $Pnma$ phase of BiFe$_{0.5}$Sc$_{0.5}$O$_3$ was found to transform into the polar $R3c$ phase identical to that of BiFeO$_3$. Subsequent cooling below the transition temperature resulted in onset of a novel polar phase with the $Ima2$ symmetry, where the ferroelectric-like displacements of Bi$^{3+}$ cations along the $[110]_p$ pseudocubic direction are combined with the antiphase octahedral tilting about the polar axis.\cite{ref:17} It has also been shown that both the $Pnma$ and the $Ima2$ polymorphs of BiFe$_{0.5}$Sc$_{0.5}$O$_3$ exhibit a long-range antiferromagnetic ordering with a weak ferromagnetic component below about 220 K. The antiferromagnetic configuration was found to be of a $G$-type (where the Fe/Sc nearest neighbors in all three directions have antiparallel spins). \\
\indent The antipolar $Pnma$ modification of BiFe$_{0.5}$Sc$_{0.5}$O$_3$ is isostructural to one of the phases of La-doped BiFeO$_3$, as reported by Rusakov et al. [\onlinecite{ref:19}] In the Bi$_{1-x}$La$_x$FeO$_3$ system, the $Pnma$ phase has been found to be stable in a narrow compositional range and an increase of the La content above $x \sim 0.19$ induced an incommensurate modulation. Based on these observations, one can expect that the antipolar modification of BiFe$_{0.5}$Sc$_{0.5}$O$_3$ can also be driven to the incommensurately modulated structure by a partial substitution of Bi with La. Indeed, our present study has revealed signs of an incommensurate modulation in the perovskite phase of the Bi$_{0.8}$La$_{0.2}$Fe$_{0.5}$Sc$_{0.5}$O$_3$ composition. In this work, comprehensive structural and magnetic studies of the incommensurate phase have been performed. The main goal was to explore how the structural modulation affects the spin ordering in the system. Data of the magnetization measurements and the neutron diffraction experiments were analyzed based on the symmetry considerations. We conclude that the magnetic structure of such a phase is a unique example, where a dominant commensurate antiferromagnetic component coexists with a macroscopic ferromagnetism and an incommensurate spin density wave that has a propagation vector related to the structural modulation. The coupling mechanism has been suggested to be the antisymmetric Dzyaloshinskii-Moriya exchange which also fully defines the spin directions in the structure.\\

\section{Experimental section}

\begin{figure}[t]
\includegraphics[scale=1.15]{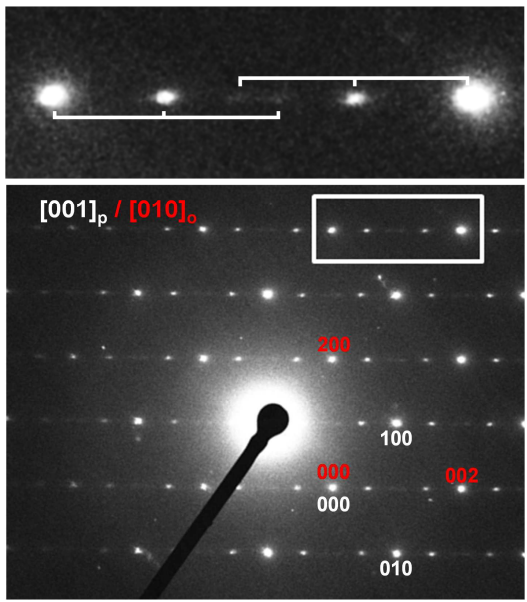}
\caption{(Color online) Electron diffraction pattern of the incommensurately modulated phase of Bi$_{0.8}$La$_{0.2}$Fe$_{0.5}$Sc$_{0.5}$O$_3$. Indexation of the fundamental spots is done in both pseudocubic (white) and the average orthorhombic (red) cells. The enlarged part of the diffraction pattern (on the top) demonstrates non-overlapping second order satellites. }
\label{fig:1}
\end{figure}
\indent High-purity oxides Bi$_2$O$_3$, La$_2$O$_3$, Fe$_2$O$_3$, and Sc$_2$O$_3$ were used as starting reagents to prepare the compositions of the Bi$_{1-x}$La$_x$Fe$_{0.5}$Sc$_{0.5}$O$_3$ series. Previously calcined oxides were mixed in the stoichiometric ratio, ball-milled in acetone, dried, and pressed into pellets. The pellets were heated in a closed alumina crucible at 1140 K for 10 min and then quenched down to room temperature. The obtained material served as a precursor for the high pressure synthesis. The pressure was generated using an anvil press DO-138A with a press capacity up to 6300 kN. In order to avoid penetration of graphite from the tubular heater to the sample a protective screen of molybdenum foil was used. The samples were synthesized at 6 GPa and 1500-1600 K. The high-pressure treatment time did not exceed 5 min.\\
\indent An X-ray diffraction study of the powdered samples was performed using a PANalytical XPert MPD PRO diffractometer (Ni-filtered Cu K$_{\alpha }$ radiation, tube power 45 kV, 40 mA; PIXEL detector, and the exposition corresponded to about 2 s per step of $0.02^{\circ }$ over the angular range $15^{\circ} - 100^{\circ }$) at room temperature.\\
\indent Electron diffraction patterns of the samples were recorded using a 200 kV JEOL 2200FS transmission electron microscope (TEM). The samples were crushed and milled with mortar and pestle. The obtained fine powder was dispersed in ethanol and deposited on a TEM grid. \\
\indent Neutron powder diffraction data for the Bi$_{0.8}$La$_{0.2}$Fe$_{0.5}$Sc$_{0.5}$O$_3$ composition were collected at the ISIS pulsed neutron and muon facility of the Rutherford Appleton Laboratory (UK), on the WISH diffractometer located at the second target
station.\cite{ref:20} The sample ($\sim$25 mg) was loaded into a cylindrical 3 mm diameter vanadium can and measured in the temperature range of 1.5 - 300 K (step 30 K, exposition time 2h) using an Oxford Instrument Cryostat. Rietveld refinements of the crystal and magnetic structures were performed using the JANA 2006 program\cite{ref:21} against the data measured in detector banks at average $2\theta$ values of 58$^{\circ }$, 90$^{\circ }$, 122$^{\circ }$, and 154$^{\circ }$, each covering 32$^{\circ }$ of the scattering plane. Group-theoretical calculations were done using ISOTROPY\cite{ref:24}, ISODISTORT\cite{ref:25} and Bilbao Crystallographic Server  software (REPRES\cite{ref:27} and Magnetic Symmetry and Applications\cite{ref:27a}). \\
\indent Magnetization data were measured using a superconducting quantum interference device (SQUID) magnetometer (Quantum Design MPMS).\\

\section{Result and discussion}

\subsection{Crystal structure}

\begin{figure*}[t]
\includegraphics[scale=1.50]{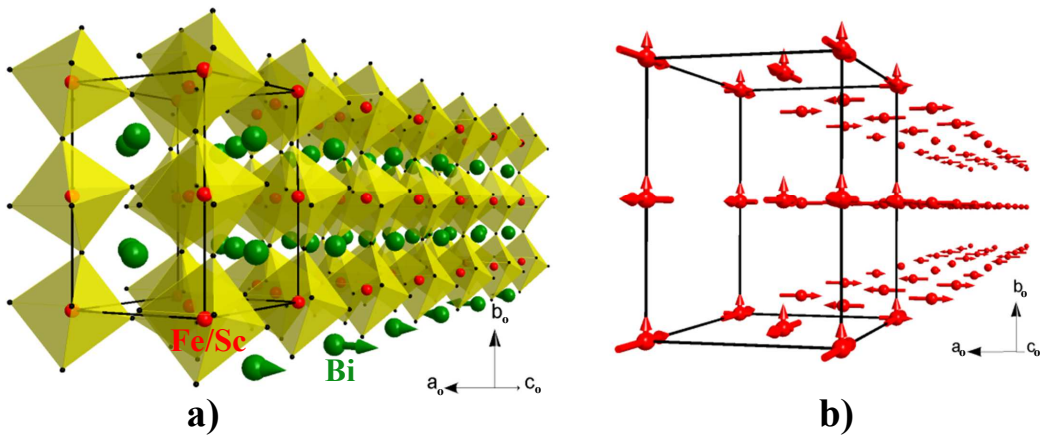}
\caption{(Color online) Crystal (a) and magnetic (b) structures of Bi$_{0.8}$La$_{0.2}$Fe$_{0.5}$Sc$_{0.5}$O$_3$. The crystal structure involves incommensurate atomic displacements (mainly Bi and oxygen as follows from the quantitative structure refinement) along the $a_o$-axis, combined with anti-phase octahedral tilting about this direction. Arrows indicate the incommensurate displacements of Bi (shown only for one Bi chain for clarity). The magnetic structure combines the primary antiferromagnetic component $G_z$ commensurate with the average structure and two secondary ones, namely the commensurate weak ferromagnetic $F_y$ and the incommensurate spin-density wave. These spin components are schematically shown as arrows along the $c_o$-, $b_o$- and $a_o$-axes, respectively.}
\label{fig:2}
\end{figure*}
\begin{table}[b]
\caption{Structural parameters of Bi$_{0.8}$La$_{0.2}$Fe$_{0.5}$Sc$_{0.5}$O$_3$ obtained from the joint refinement of room-temperature X-ray and neutron diffraction data using the $Imma(0,0,\gamma )s00$ superspace group with the basis vectors related to the parent cubic $Pm\bar{3}m$ structure as $(1,-1,0,0),(0,0,-2,0),(1,1,0,0),(0,0,0,1)$ and origin at $(0,1,1,0)$. Unit cell parameters $a_o=5.6879(1)\AA$, $b_o=7.9593(1)\AA$, $c_o=5.7166(1)\AA$, $\bm {k}_o^{\Lambda }=\gamma \bm {c}^*_o=0.534(3) \bm {c}^*_o$. Reliability factors $R_p=2.79\%$ and $R_{wp}=3.25\%$.}
\centering 
\begin{tabular*}{0.48\textwidth}{@{\extracolsep{\fill}} l c c c c} 
\hline\hline\\ [-1.5ex] 
 Atom  & $x$ & $y$ & $z$ & $U_{iso}$\\ 
 $A^1_i${\footnotemark[1]}  & $A^1_x$ & $A^1_y$ & $A^1_z$ & \\
  $B^1_i${\footnotemark[1]}  & $B^1_x$ & $B^1_y$ & $B^1_z$ & \\
\hline\\ [-1.5ex] 
Bi/La & 0 & 0.25 & 0.5093(4) & 0.024(1) \\ 
 & 0.0475(7) & 0 & 0 &  \\
 & -0.033(1) & 0 & 0 &  \\
Fe & 0 & 0 & 0 & 0.010(3)\\
 & 0 & 0 & 0 & \\
 & 0 & 0 & 0 & \\
O1 & 0.25 & -0.0444(1) & 0.25 & 0.037(1)\\
 & 0.023(1) & 0 & 0.012(2) & \\
 & -0.0043(7) & 0 & 0 & \\
O2 & 0 & 0.25 & 0.0785(4) & 0.01(1)\\
 & -0.0929(8) & 0 & 0 & \\
 & -0.005(2) & 0 & 0 & \\
\hline
\hline  
\end{tabular*}
\footnotetext[1]{$A^1_i$ and $B^1_i$ ($i=x,y,z$) are the Fourier coefficients of the first harmonic ($n=1$) of the displacive modulation function: $u_{i,j,l}(\bm {r}_{j,l}\cdot \bm {k}_o^{\Lambda } )=\sum_{n=0}^{\infty} A^n_{i,j} sin (2\pi n [\bm {r}_{j,l}\cdot \bm {k}_o^{\Lambda }]) + $ \\  $B^n_{i,j} cos (2\pi n [\bm {r}_{j,l}\cdot \bm {k}_o^{\Lambda }])$, where $\bm {r}_{j,l}$ indicates the position of the $j$-th atom of the average structure in the $l$-th unit cell.}
\label{tab:1} 
\end{table}
\indent It was found from the obtained room temperature X-ray and neutron diffraction data that the crystal structure of Bi$_{0.8}$La$_{0.2}$Fe$_{0.5}$Sc$_{0.5}$O$_3$  is different from the orthorhombic $Pnma$ structure of undoped BiFe$_{0.5}$Sc$_{0.5}$O$_3$, indicating a compositionally-driven phase transition. The indexation procedure of the powder diffraction patterns appeared to be difficult using a reasonable size superstructure. Taking into account that about the same concentration of La has been reported by Rusakov et al. [\onlinecite{ref:19}] to induce an incommensurate phase in the Bi$_{1-x}$La$_x$FeO$_3$ system, an electron diffraction on Bi$_{0.8}$La$_{0.2}$Fe$_{0.5}$Sc$_{0.5}$O$_3$ has been performed. The measurements confirmed the presence of an incommensurate modulation with the propagation vector $\bm {k}_p^{\Sigma }=(\alpha,\alpha,0;\alpha \sim 0.27)$ with respect to the cubic perovskite unit cell (Fig. \ref{fig:1}). This propagation vector is the $\Sigma $-line of symmetry, following the ISOTROPY notations,\cite{ref:24,ref:25} and the associated distortion is present in the antiferroelectric structures of the undoped BiFe$_{0.5}$Sc$_{0.5}$O$_3$ as well as in the closely related $Pbma$ structure of PbZrO$_3$ ($\Sigma_2$-distortions).\cite{ref:22} In both cases, the parameter $\alpha $ takes the commensurate value of 1/4 and therefore these antiferroelectric structures can be regarded as the lock-in phases. Since the value of 1/4 is not stimulated by symmetry, it can be changed by either external perturbations or changes in chemical composition. This important symmetry aspect has been recently highlighted by Tagantsev et al. in the study of lattice dynamics of PbZrO$_3$.\cite{ref:23} These authors concluded that the antiferroelectric state is a 'missed' incommensurate phase and that the transition to this state is driven by softening of a single polar lattice mode. Due to flexoelectric coupling, the system is expected to be virtually unstable against the incommensurate modulation as was shown by Axe et al.\cite{ref:23a} However, the Umklapp interaction forces the system to go directly to the commensurate lock-in phase, leaving the incommensurate phase as a 'missed' opportunity.\cite{ref:23}\\
\begin{figure}[t]
\includegraphics[scale=1.16]{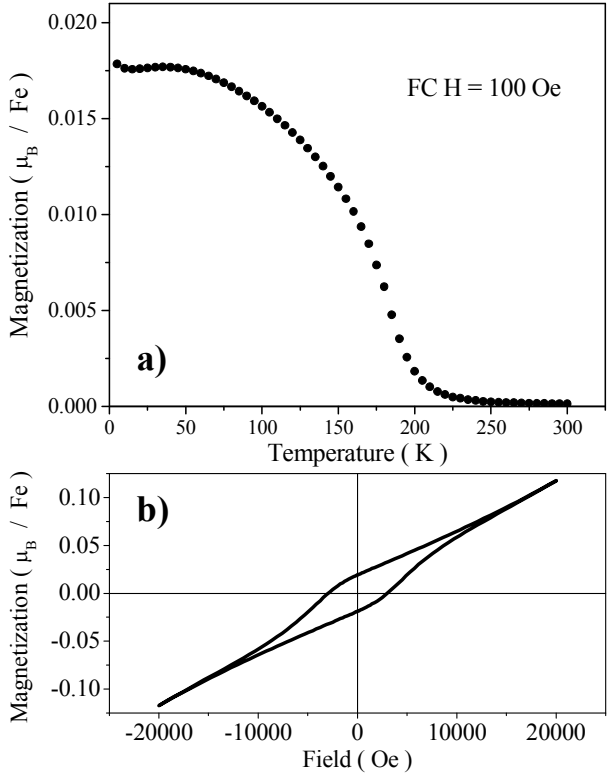}
\caption{Magnetization as a function of temperature, measured for Bi$_{0.8}$La$_{0.2}$Fe$_{0.5}$Sc$_{0.5}$O$_3$ under the magnetic field of H=100 Oe after cooling in this field (a). Magnetization loop measured at 5 K after a zero-field cooling (b).}
\label{fig:3}
\end{figure}
\begin{figure}[t]
\includegraphics[scale=0.7]{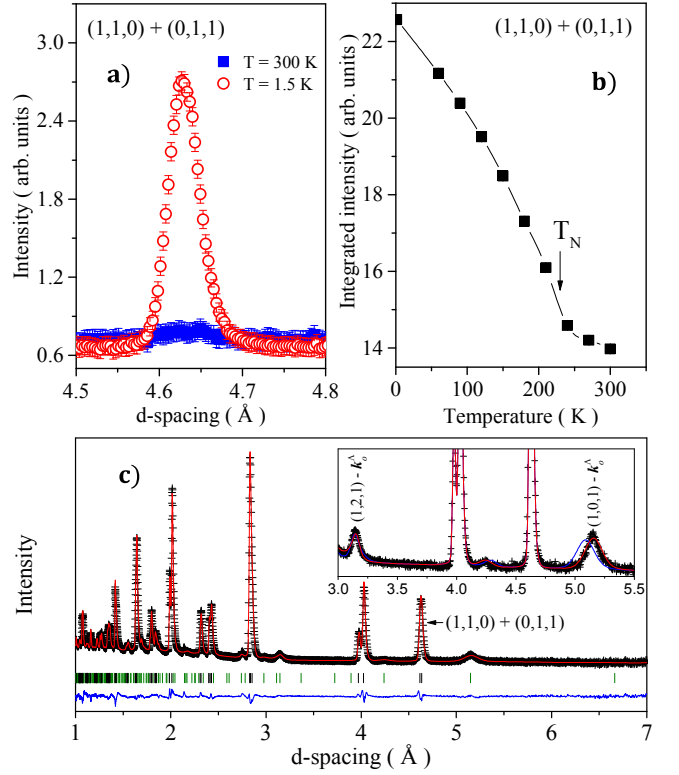}
\caption{(Color online) (a) Neutron diffraction patterns of Bi$_{0.8}$La$_{0.2}$Fe$_{0.5}$Sc$_{0.5}$O$_3$ at the vicinity of the strongest magnetic peaks collected above and below T$_N$. (b) Integrated intensity of the magnetic peaks as a function of temperature (error bars are smaller than the size of the symbol). (c) Rietveld refinement of the neutron diffraction data collected at 1.5 K. The cross symbols and solid line (red) represent the experimental and calculated in the $Im'ma'(0,0,\gamma )s00$ magnetic superspace group intensities, respectively, and the line below (blue) is the difference between them. Tick marks indicate the positions of Bragg peaks (green for satellites and black for fundamental). Inset shows an enlarged part of the diffraction pattern where the strongest structural satellites are observed. The solid red and blue lines represent intensities calculated for the incommensurate ($\gamma =0.534(3)$) and commensurate ($\gamma=0.5$) values of the modulation vector $\bm {k}_o^{\Lambda }$, respectively. }
\label{fig:4}
\end{figure}
\indent This consideration provides a way to deduce the appropriate symmetry of the modulated phase of Bi$_{0.8}$La$_{0.2}$Fe$_{0.5}$Sc$_{0.5}$O$_3$. Our experimental diffraction data indicate the presence of only $R$- and $\Sigma $-type superstructures which results in a few (3+1) superspace isotropy subgroups to be tested in the refinement procedure.\cite{ref:24,ref:25} A combination of the single-$\bm {k}_p^{\Sigma }$, $\Sigma_2 $  incommensurate modulation with anti-phase octahedral tilting ($R^+_4$ commensurate distortions with the $\bm {k}_p^R = (\frac{1}{2},\frac{1}{2},\frac{1}{2})$ propagation vector) results in seven distinct superspace subgroups\cite{ref:24,ref:25} but only three of them are consistent with the orthorhombic metric of the pseudocubic perovskite unit cell. The quantitative joint refinement of the X-ray and neutron diffraction patterns confirmed the $Imma(0,0,\gamma )s00$ superspace group to be the adequate one to describe the modulated structure of Bi$_{0.8}$La$_{0.2}$Fe$_{0.5}$Sc$_{0.5}$O$_3$. The corresponding coupled order parameter takes the $(0,\delta ,-\delta;\eta,0,0,0,0,0,0,0,0,0,0,0)$ direction in the reducible $R^+_4\oplus \Sigma_2$ representation space. The structure involves incommensurate atomic displacements (predominantly Bi and oxygen) along the $[1 \bar{1} 0]_p$ pseudocubic direction combined with anti-phase octahedral tilting about this axis (Fig. \ref{fig:2}(a)). This structural model is identical to the 'missed' incommensurate phase discussed by Tagantsev et al. [\onlinecite{ref:23}] for the antiferroelectric PbZrO$_3$. Earlier, Rusakov et al. [\onlinecite{ref:19}] used the $Imma(0,0,\gamma )s00$ superspace group to propose a structural model for the incommensurate phase of Bi$_{0.75}$La$_{0.25}$FeO$_3$. The authors split the Bi position and applied a step-like occupational modulation to model the constant and correlated shifts of Bi and oxygen atoms along the $a_o$-axis of the average $Imma$ structure. Hereafter we use $o$-subscript to denote the orthorombic setting (see Table \ref{tab:1} for the relation between the orthorhombic and cubic settings). In our refinement, the modulation was applied to the atomic displacements. In such an approach, the modulation is a characteristic of the displacive correlation function averaged over the sample volume (Table \ref{tab:1}). This model is hardly appropriate to analyze the local bond distances and angles but sufficient to consider the symmetry-controlled physical properties; in particular, a coupling between the orthogonal magnetic modes. It should be pointed out that although the model proposed by Rusakov et al. [\onlinecite{ref:19}] is more adequate to discuss the local crystal chemistry, it is another limited case which ignores the fact that the amplitude of the atomic displacements depends on the local fluctuations of La.\\

\subsection{Magnetic structure}

\indent Magnetization measurements of Bi$_{0.8}$La$_{0.2}$Fe$_{0.5}$Sc$_{0.5}$O$_3$ revealed a weak ferromagnetic behavior below $T_\textrm{N} \sim 220$ K (Fig. \ref{fig:3}(a)) with the value of the spontaneous moment of $\sim 0.022 \mu_B$/Fe (Fig. \ref{fig:3}(b)). This value as well as the critical temperature of the magnetic ordering are close to those in the polar and antipolar polymorphs of BiFe$_{0.5}$Sc$_{0.5}$O$_3$.\cite{ref:17} In agreement with the magnetization data, the neutron diffraction measurements indicate a long-range antiferromagnetic ordering below $T_\textrm{N}$ (Fig. \ref{fig:4}(a,b)). Note, a precise determination of the critical temperature from the neutron diffraction data is complicated by a superposition of the magnetic Bragg intensity with a diffuse component developing above T$_N$. The diffuse component is probably related to a magnetic inhomogeneity caused by local fluctuations in the Fe/Sc ratio. The magnetic Bragg reflections, observed below T$_N$, are resolution limited and can be indexed using the average orthorhombic $Imma$ structure assuming $\bm {k}_o^{m\Gamma }=0$ propagation vector. To refine quantitatively the magnetic structure, we classified the magnetic modes according to the time-odd irreducible representations of the $Imma$ space group. The crystal structure based on which the magnetic order emerges is however incommensurate. Therefore, one needs to decompose the primary structural modulation in respect of the time-even irreducible representations of $Imma$. The relevant analysis revealed that the incommensurate structural modulation has the symmetry of the $\Lambda_4$ representation associated with the $\bm {k}_o^{\Lambda }=(0,0,\gamma )$ line of symmetry.\cite{ref:25,ref:25} Then, by combining the time-odd $\bm {k}_o^{m\Gamma }=0$ representations with the $\Lambda_4$ incommensurate order parameter, we deduced the magnetic superspace groups\cite{ref:25,ref:25} and tested them in the refinement versus the neutron diffraction data. The $m\Gamma^+_4\oplus \Lambda_4$ combination resulting in the $Im'ma'(0,0,\gamma)s00$ magnetic superspace group was found to provide a uniquely good refinement quality (Fig. \ref{fig:4}(c)). Remarkably, this magnetic superspace group implies a coexistence of both commensurate and incommensurate components of the magnetic order parameter (Fig. \ref{fig:2}(b)). The former is represented by ferromagnetic ($F$) and $G$-type antiferromagnetic components along the $b_o$- and $c_o$-axis of the average $Imma$ structure, respectively. The latter is along the $a_o$-axis, with the modulation related to the structural one by the propagation vector conservation law as will be discussed below. The refinement procedure yielded the statistically significant value for the commensurate antiferromagnetic component only: $3.90(4)\mu_B$ per Fe. This value is somewhat smaller than the expected one, 5 $\mu_B$, for the $e^2_gt^3_{2g}$ electronic configuration of Fe$^{3+}$ but practically identical to the ordered moment in Bi$_{0.9}$La$_{0.1}$FeO$_3$.\cite{ref:23b} The ferromagnetic component in Bi$_{0.8}$La$_{0.2}$Fe$_{0.5}$Sc$_{0.5}$O$_3$ is clearly evidenced by the magnetization data (Fig. \ref{fig:3}(b)) with the value well beyond the capabilities of the unpolarised neutron powder diffraction experiment. Although the presence of the modulated component does not follow directly from the neutron and magnetization data, it can be shown that the system couples it to gain full advantage of the antisymmetric exchange interactions. These interactions are anisotropic and force the direction of the primary $G$-type antiferromagnetic component to be along the $c_o$-axis.\\
\indent The Heisenberg symmetric exchange interactions are degenerate in respect of the global spin rotations and therefore the corresponding part of the exchange energy does not depend on the crystallographic direction of the interacting spins. These interactions usually dominate and, in a first approximation, define the relative orientations of spins in a magnetic structure. Then, the higher order anisotropic terms in the magnetic Hamiltonian such as antisymmetric exchange, single ion anisotropy, magnetoelastic coupling and dipole-dipole interactions remove the global spin rotation degeneracy, making some crystallographic directions preferable. In the Bi$_{0.8}$La$_{0.2}$Fe$_{0.5}$Sc$_{0.5}$O$_3$ perovskite, the octahedrally coordinated Fe$^{3+}$ ions with a nearly quenched orbital degree of freedom ($L=0, S=5/2$) interact antiferromagnetically via the strong superexchange, which results in the $G$-type magnetic structure as found experimentally and as observed in many other Fe$^{3+}$-based perovskites.\cite{ref:26} Structural distortions activate the anisotropic terms; in particular, antisymmetric exchange.\\ 
\indent As follows from the structure refinement, described in the previous section, there are two primary structural distortions in Bi$_{0.8}$La$_{0.2}$Fe$_{0.5}$Sc$_{0.5}$O$_3$, namely anti-phase octahedral tilting and incommensurately modulated Bi and oxygen displacements. The part of the antisymmetric exchange related to octahedral tilting was considered in details in Ref. [\onlinecite{ref:17}]. It has been shown that the axial distortions associated with the octahedral rotations are responsible for the weak ferromagnetic properties of antiferromagnetically ordered perovskites with a $G$-type spin configuration. The relevant part of the Dzyaloshinskii vector, $\bm {D}^{oct}_{i,j}$, is expressed by the antiferroaxial vector which is a characteristic of the tilting pattern. In other words, the component of the Dzyaloshinskii vector which induces the weak ferromagnetism coincides with the tilting axis of octahedra. One can therefore expect that the spin components of the primary $G$-type antiferromagnetic mode are confined within the ($b_oc_o$) plane, since the octahedra are tilted about the $a_o$-axis in the average $Imma$ structure. This part of the antisymmetric exchange energy, $\bm {D}^{oct}_{i,j}\cdot [\bm {S}_{i}\times \bm {S}_{j}]$, is degenerate in respect of the moment direction in the ($b_oc_o$) plane. This degeneracy is however removed, when we take into account the antisymmetric exchange imposed by the incommensurate structural distortion.\\
\indent To demonstrate that, we need to work out the appropriate free-energy coupling terms in respect of the parent $Pm \bar{3}m$ symmetry (see Appendix for details). The incommensurate structural distortion transforms as the twelve-dimensional $\Sigma_2(\eta_{j=1-12})$ irreducible representation associated with the $\bm {k}_p^{\Sigma }=(\alpha,\alpha ,0 )$ propagation vector of the cubic space group $(\alpha=\frac{\gamma }{2})$. The direction of the order parameter in the $\Sigma_2$ representation space is specified by the single non-zero component $\eta_1$. The symmetry properties of the $G$-type antiferromagnetic mode are defined by the $\bm {k}_p^{mR}=(\frac{1}{2},\frac{1}{2},\frac{1}{2})$ propagation vector and the three-dimensional representation $mR^+_4(\mu_{i=1-3})$. A combination of these order parameters to a free-energy invariant requires a coupling to a time-odd (magnetic) physical quantity, $\xi $, with the modulation related to the structural one as $\bm {k}_p^{mS}=(\frac{1}{2}-\alpha,\frac{1}{2}-\alpha,\frac{1}{2})$ to maintain the translational invariance. As detailed in the Appendix, a third power invariant with a magnetic dipole order parameter can be formed only when the spins in the primary $G$-type antiferromagnetic mode are along the orthorhombic $c_o$-axis $(\mu_1=0, \mu_2=\mu_3 \neq 0)$. The relevant energy term is:
\begin{eqnarray}
\mu_2 \eta_1 \xi_{10} + \mu_3 \eta_1 \xi_{10} \equiv 2\mu \eta \xi
\label{eq:1}
\end{eqnarray}
which describes a coupling of the spin density wave with the spin components being along the $a_o$-axis of the average $Imma$ structure (Fig. \ref{fig:2}(b)). The presence of this incommensurately modulated spin component is in full agreement with the magnetic $Im'ma'(0,0,\gamma )s00$ superspace symmetry derived above. In the framework of the representation theory, the symmetry properties of the spin density wave are specified by the twelve-dimensional time-odd irreducible representation $mS_3(\xi_{l=1-12})$  with the single non-zero component of the order parameter $\xi_{10}$. The crucial point is that there is no such coupling, if the spins in the antiferromagnetic configuration are along the $b_o$-axis $(\mu_1 \neq 0, \mu_2=\mu_3 = 0)$. Thus, the system chooses the $c_o$-axis for the spins direction in the primary antiferromagnetic $G$-mode to activate the energy term specified by expression (\ref{eq:1}), which breaks the degeneracy between the $b_o$- and $c_o$-axes.\\ 
\indent The free-energy term specified by expression (\ref{eq:1}) is not invariant under a global spin rotation (it vanishes when the spins are along the orthorhombic $b_o$-axis) revealing its relativistic nature.\cite{ref:28} This invariant implies that the incommensurate atomic displacements modulate the relevant component of the Dzyaloshinskii vector which in turn induces the spin density wave through the relativistic antisymmetric exchange.\\
\indent Thus, the magnetic structure of Bi$_{0.8}$La$_{0.2}$Fe$_{0.5}$Sc$_{0.5}$O$_3$ can be fully understood by taking into account only the isotropic symmetric and anisotropic antisymmetric exchange interactions. The first type of interactions defines the primary $G$-type antiferromagnetic configuration through the strong 180-degree superexchange expected for the half-occupied $e_g$ orbitals of Fe$^{3+}$. The second (anisotropic) type of interactions chooses the $c_o$-axis for the spins direction in the primary mode to fully exploit the dominant structural distortions and couple the secondary orthogonal ferromagnetic and incommensurate spin components.\\

\section{Conclusions}

\indent A 20\% substitution of Bi with La in the Bi$_{1-x}$La$_x$Fe$_{0.5}$Sc$_{0.5}$O$_3$ system synthetized under high-pressure and high-temperature conditions induces incommensurately modulated structural phase. The symmetry of this phase is described by the $Imma(0,0,\gamma )s00$ superspace group ($\gamma =0.534(3)$) with modulated displacements of Bi/La and oxygen ions. The structure combines the same type of primary distortions, as the prototype antiferroelectric structure of PbZrO$_3$. The difference between the two structures is in the value of the propagation vector for the antipolar displacements ($\Sigma_2$ distortive mode) which is commensurate ($\gamma =0.5$) in the case of PbZrO$_3$. Both propagation vectors (commensurate and incommensurate), however, belong to the same line of symmetry and therefore the commensurate value (lock-in phase) is not stimulated by the symmetry and can be tuned by composition or external perturbations such as pressure, strain (for thin-film forms) and electric field.\\
\indent Below $T_\textrm{N} \sim220$ K, a long range antiferromagnetic ordering commensurate with the average $Imma$ structure takes place. The spins are aligned along the $c_o$-axis which allows the system to gain energy from the antisymmetric exchange activated by the two primary structural distortions: namely, octahedral tilting and the incommensurate atomic displacements. The antisymmetric exchange imposed by the tilting induces a weak ferromagnetic component along the $b_o$-axis. The part of the antisymmetric exchange related to the modulated atomic displacements with $\bm {k}_p^{\Sigma }=(\alpha, \alpha, 0)$ couples a spin-density wave with the propagation vector $\bm {k}_p^{mS }=(\frac{1}{2}-\alpha, \frac{1}{2}-\alpha, \frac{1}{2})$ and the spin components being along the $a_o$-axis. These results demonstrate the crucial role of the antisymmetric exchange in magnetic properties of Fe$^{3+}$-containing distorted perovskites. \\

\section*{Acknowledgement}
 
\indent This work was supported by project TUMOCS. This project has received funding from the European Union's Horizon 2020 research and innovation programme under the Marie Sk\l{}odowska-Curie grant agreement No 645660. 

\section*{Appendix}

\indent To explore the Landau free-energy terms activated by the incommensurate structural modulation, let us define the transformational properties of the distortions involved in respect of the parent $Pm \bar{3}m$ symmetry. In this cubic structure, Bi/La, Fe/Sc and oxygen atoms occupy $1b(\frac{1}{2},\frac{1}{2},\frac{1}{2})$, $1a(0,0,0)$ and $3d(\frac{1}{2},0,0)$ Wyckoff positions, respectively. The structural modulation and the $G$-type spin configuration transform as the twelve-dimensional time-even $\Sigma_2(\eta_1,\eta_2,\eta_3,\eta_4,\eta_5,\eta_6,\eta_7,\eta_8,\eta_9,\eta_{10},\eta_{11},\eta_{12})$, $\{ \bm {k}_p^{\Sigma }=(\alpha ,\alpha ,0) \}$ and the three-dimensional time-odd $mR^+_4(\mu_1,\mu_2,\mu_3)$, $\{ \bm {k}^{mR}_p=(\frac{1}{2},\frac{1}{2},\frac{1}{2}) \}$ irreducible representations of the cubic space group ($\{ \}$ indicates a wave vector star with a representative arm enclosed).\cite{ref:24,ref:25} The lowest degree coupling term maintaining the time reversal symmetry is a cubic trilinear invariant of the $\sum_{i,j,l} \mu_i \eta _j \xi _l$ - type. Here, $\xi_l$-denotes the coupled time-odd order parameter whose symmetry we need to figure out for the cases when the magnetic moments are along the orthorhombic $b_o$ (pseudocubic $[001]_p$) and $c_o$ (pseudocubic $[110]_p$) axes. These spin configurations occur when the $mR^+_4$ order parameter takes the $(\mu_1,0,0)$ and $(0,\mu_2,\mu_3; \mu_2=\mu_3$) directions, respectively. The incommensurate atomic displacements along the $[1\bar{1}0]_p$ pseudocubic direction (orthorhombic $a_o$-axis) are described by the $(\eta_1,0,0,0,0,0,0,0,0,0,0,0)$ order parameter in the $\Sigma_2$ representation space. The translation symmetry requires the $\xi $-quantity to be associated with the $\{ \bm {k}^{mS}_p=(\frac{1}{2}-\alpha ,\frac{1}{2}-\alpha,\frac{1}{2}) \}$ propagation vector star ($S$-line of symmetry), where $\alpha =\gamma /2$ is the wave number of the structural modulation. This comes directly from the trilinear form of the invariant which requires the product of the Fourier transforms $e^{-2 \pi i(\bm {k}_p^{\Sigma } \cdot \bm {t})}$ and $e^{2 \pi i(\bm {k}_p^{mS} \cdot \bm {t})}$ associated with the $\eta_j$ and $\xi_l$ order parameters to change sign at the ${\bm t}_{p1}=(1,0,0)$, ${\bm t}_{p2}=(1,0,0)$ and ${\bm t}_{p3}=(1,0,0)$ translations. There are four twelve-dimensional irreducible representations, $mS_{\nu =1-4}(\xi_1,\xi_2,\xi_3,\xi_4,\xi_5,\xi_6,\xi_7,\xi_8,\xi_9,\xi_{10},\xi_{11},\xi_{12})$, associated with the $\{ \bm {k}_p^{mS} \}$ star, but only three of them appear in the decomposition of the pseudovector (magnetic) reducible representation localized on the Fe $1b$ Wyckoff position:
\begin{eqnarray}
\Gamma_{mag}(1b) = mS_2 \oplus mS_3 \oplus mS_4
\label{eq:2}
\end{eqnarray}
Using the ISOTROPY software (irrep version 2011),\cite{ref:24,ref:25} one can derive that the order parameter with the $mS_1$ symmetry is coupled to the $mR^+_4\otimes \Sigma_2$ product through the general free-energy invariant of the form:
\begin{eqnarray}
\mu_{1} \eta_{1} \xi_{9} - \mu_{1} \eta_{2} \xi_{10} - \mu_{1} \eta_{3} \xi_{11} + \mu_{1} \eta_{4} \xi_{12} + \nonumber \\
\mu_{2} \eta_{9} \xi_{5}-\mu_{2} \eta_{10} \xi_{6} - \mu_{2} \eta_{11} \xi_{7} + \mu_{2} \eta_{12} \xi_{8} + \nonumber \\
\mu_{3} \eta_{5} \xi_{1} - \mu_{3} \eta_{6} \xi_{2} - \mu_{3} \eta_{7} \xi_{3} + \mu_{3} \eta_{8} \xi_{4}
\label{eq:3}
\end{eqnarray}
which is reduced down to the simple $\mu_{1} \eta_{1} \xi_{9}$ term for the relevant direction of the $\Sigma_2$ order parameter. This term describes the allowed coupling scheme for the case of the $G$-type spin configuration with the moments being along the orthorhombic $b_o$-axis (pseudocubic $[001]_p$). The $mS_1$ irreducible representation, however, does not enter into the decomposing of the reducible pseudovector representation on the Fe Wyckoff position (zero subduction frequency, see expression (\ref{eq:2})). This means that there are no any dipole magnetic modes localized on the Fe site with the symmetry of the $mS_1$ representation and therefore the system cannot activate the antisymmetric exchange through this type of the tri-linear invariant.\\
\indent The situation is different when the coupled order parameter $\xi_l$ transforms as the $mS_3$ irreducible representation. The general coupling invariant takes the form:
\begin{eqnarray}
\mu_{1} \eta_{5} \xi_{2} + \mu_{1} \eta_{6} \xi_{1} - \mu_{1} \eta_{7} \xi_{4} - \mu_{1} \eta_{8} \xi_{3} + \nonumber \\
\mu_{1} \eta_{9} \xi_{6} + \mu_{1} \eta_{10} \xi_{5} - \mu_{1} \eta_{11} \xi_{8} - \mu_{1} \eta_{12} \xi_{7} + \nonumber \\
\mu_{2} \eta_{1} \xi_{10} + \mu_{2} \eta_{2} \xi_{9} + \mu_{2} \eta_{3} \xi_{12} + \mu_{2} \eta_{4} \xi_{11} + \nonumber \\
\mu_{2} \eta_{5} \xi_{2} + \mu_{2} \eta_{6} \xi_{1} + \mu_{2} \eta_{7} \xi_{4} + \mu_{2} \eta_{8} \xi_{3} + \nonumber \\
\mu_{3} \eta_{1} \xi_{10} + \mu_{3} \eta_{2} \xi_{9} - \mu_{3} \eta_{3} \xi_{12} - \mu_{3} \eta_{4} \xi_{11} + \nonumber \\
\mu_{3} \eta_{9} \xi_{6} + \mu_{3} \eta_{10} \xi_{5} + \mu_{3} \eta_{11} \xi_{8} + \mu_{3} \eta_{12} \xi_{7}
\label{eq:4}
\end{eqnarray}
with the non-vanishing terms $\mu_{2} \eta_{1} \xi_{10} + \mu_{3} \eta_{1} \xi_{10}$ for the $(\eta_1,0,0,0,0,0,0,0,0,0,0,0)$ direction describing the incommensurate structural modulation. These terms are activated when the spins of the primary $G$-type antiferromagnetic configuration are along the orthorhombic $c_o$-axis (pseudocubic $[110]_p$). The subduction frequency of the $mS_3$ representation is non-zero for the $1b$ pseudovector reducible representation ex.(\ref{eq:2}) and the projection operator yields the spin density wave localized on the Fe site with the spin direction along the orthorhombic $a_o$-axis (pseudocubic $[1\bar{1}0]_p$) (Fig. \ref{fig:2}b).\\
\indent Finally, let us point out, for completeness, that the magnetic order parameter transforming as the $mS_2$ irreducible representation does not form a third power invariant with the $mR^+_4 \otimes \Sigma_2$ product. The $mS_4$ representation does form such invariant and provides coupling, when the spins, in the primary $G$-type configuration, are along the orthorhombic $a_o$-axis. The latter situation is however unfavourable for the part of the antisymmetric exchange associated with the octahedral tilting as discussed in the main text.


\begin{thebibliography}{99}\label{sec:TeXbooks}
\bibitem{ref:1} G. Catalan, J. F. Scott, Adv. Mater. {\bf{21}}, 2463 (2009).
\bibitem{ref:2} A. P. Pyatakov, A. K. Zvezdin,  Physics - Uspekhi {\bf{55}}, 557 (2012).
\bibitem{ref:3} Y. Shimakawa, M. Azuma and N. Ichikawa, Materials {\bf{4}}, 153 (2009).
\bibitem{ref:4} M. Azuma, K. Takata, T. Saito, S. Ishiwata, Y. Shimakawa, M. Takano, J. Am. Chem. Soc. {\bf{127}}, 8889 (2005).
\bibitem{ref:5} V. A. Khomchenko, D. A. Kiselev, M. Kopcewicz, M. Maglione, V. V. Shvartsman, P. Borisov, W. Kleemann, A. M. L. Lopes, Y. G. Pogorelov, J. P. Araujo, R. M. Rubinger, N. A. Sobolev, J. M. Vieira, A. L. Kholkin, J. Magn. Magn. Mater. {\bf{321}}, 1692 (2009).
\bibitem{ref:6} V. A. Khomchenko, D. A. Kiselev, I. K. Bdikin, V. V. Shvartsman, P. Borisov, W. Kleemann, J. M. Vieira and A. L. Kholkin, Appl. Phys. Lett. {\bf{93}}, 262905 (2008).
\bibitem{ref:7} S. Karimi, I. M. Reaney, I. Levin, I. Sterianou,  Appl. Phys. Lett. {\bf{94}}, 112903 (2009).
\bibitem{ref:8} I. Levin, M. G. Tucker, H. Wu, V. Provenzano, C. L. Dennis, S. Karimi, T. Comyn, T. Stevenson, R. I. Smith, I. M.  Reaney, Chem. Mater. {\bf{23}}, 2166 (2011). 
\bibitem{ref:9} I. O. Troyanchuk, D. V. Karpinsky, M. V. Bushinsky, V. A. Khomchenko, G. N. Kakazei, J. P. Araujo, M. Tovar, V. Sikolenko, V. Efimov, and A. L. Kholkin, Phys. Rev. B {\bf{83}}, 054109 (2011).
\bibitem{ref:10} I. O. Troyanchuk, D. V. Karpinsky, M. V. Bushinsky, O. S. Mantytskaya, N. V. Tereshko, and V. N. Shut, J. Am. Ceram. Soc. {\bf{94}}, 4502 (2011).
\bibitem{ref:11} D. V. Karpinsky, I. O. Troyanchuk, J. V. Vidal, N. A. Sobolev, A. L. Kholkin, Solid State Commun. {\bf{151}}, 536 (2011)
\bibitem{ref:12} A. A Belik, A. M. Abakumov, A. A. Tsirlin, J. Hadermann, J. Kim, G. Van Tendeloo, E. Takayama-Muromachi,  Chem. Mater. {\bf{23}}, 4505 (2011).
\bibitem{ref:13} D. P. Kozlenko, A. A. Belik, A. V. Belushkin, E. V. Lukin, W. G. Marshall, B. N. Savenko, and E. Takayama-Muromachi, Phys. Rev. B {\bf{84}}, 094108 (2011).
\bibitem{ref:14} S. Prosandeev, D. Wang, W. Ren, J. Iniguez, L. Bellaiche,  Adv. Funct. Mater. {\bf{23}}, 234 (2013).
\bibitem{ref:15} A. A. Belik, H. Yusa, N. Hirao, Y. Ohishi and E. Takayama-Muromachi, Chem. Mater. {\bf{21}}, 3400 (2009). 
\bibitem{ref:16} C. S. Knee, M. G. Tucker, P. Manuel, S. Cai, J. Bielecki, L. Börjesson, and S. G. Eriksson, Chem. Mater. {\bf{26}}, 1180 (2014).
\bibitem{ref:17} D. D. Khalyavin, A. N. Salak, N. M. Olekhnovich, A. V. Pushkarev, Yu. V. Radyush, P. Manuel, I. P. Raevski, M. L. Zheludkevich, and M. G. S. Ferreira, Phys. Rev. B {\bf{89}}, 174414 (2014).
\bibitem{ref:18} S A. Prosandeev, D. D. Khalyavin, I. P. Raevski, A. N. Salak, N. M. Olekhnovich, A. V. Pushkarev, and Yu V. Radyush, Phys. Rev. B {\bf{90}}, 054110 (2014).
\bibitem{ref:19} D. A. Rusakov,  A. M. Abakumov, K. Yamaura, A. A. Belik, G. Van Tendeloo, E. Takayama-Muromachi,  Chem. Mater. {\bf{23}}, 285 (2011).
\bibitem{ref:20} L. C. Chapon, P. Manuel, P. G. Radaelli, C. Benson, L. Perrott, S. Ansell, N. J. Rhodes, D. Raspino, D. Duxbury, E. Spill, J. Norris,  Neutron News {\bf{22}}, 22 (2011).
\bibitem{ref:21} V. Petricek, M. Dusek, and L. Palatinus,  Z. Kristallogr. {\bf{229}}, 345 (2014).
\bibitem{ref:24} H. T. Stokes, D. M. Hatch, and B. J. Campbell, ISOTROPY Software Suite, iso.byu.edu.
\bibitem{ref:25} B. J. Campbell, H. T. Stokes, D. E. Tanner, and D. M. Hatch, J. Appl. Crystallogr. {\bf{39}}, 607 (2006).
\bibitem{ref:27} M. I. Aroyo, A. Kirov, C. Capillas, J. M. Perez-Mato and H. Wondratschek, Acta Cryst. {\bf{A62}}, 115 (2006).
\bibitem{ref:27a} J. M. Perez-Mato, S. V. Gallego, E. S. Tasci, L. Elcoro, G. de la Flor, and M.I. Aroyo, Annu. Rev. Mater. Res. {\bf{45}}, 217 (2015).
\bibitem{ref:22} D. D. Khalyavin, A. N. Salak, N. P. Vyshatko, A. B. Lopes, N. M. Olekhnovich, A. V. Pushkarev, I. I. Maroz, Yu. V. Radyush,  Chem. Mater. {\bf{18}}, 5104 (2006).
\bibitem{ref:23} A. K. Tagantsev, K. Vaideeswaran, S. B. Vakhrushev, A. V. Filimonov, R. G. Burkovsky, A. Shaganov,
D. Andronikova, A. I. Rudskoy, A. Q. R. Baron, H. Uchiyama, D. Chernyshov, A. Bosak, Z. Ujma, K. Roleder, A. Majchrowski, J.-H. Ko and N. Setter, Nat. Commun. {\bf{4}}, 2229 (2013).
\bibitem{ref:23a} J. D. Axe, J. Harada, and G. Shirane, Phys. Rev. B {\bf{1}}, 1227 (1970).
\bibitem{ref:23b} C. S. Knee, M. G. Tucker, P. Manuel, S. Cai, J. Bielecki, L. Borjesson, and S. G. Eriksson, Chem. Mater. {\bf{26}}, 1180 (2014).
\bibitem{ref:26} Yu. A. Izumov, V. E. Naish, and R. P. Ozerov, {\it Neutron Diffraction of Magnetic Materials} (Consulting Bureau, New York, 1991). 
\bibitem{ref:28} I. E. Dzyaloshinskii, Sov. Phys. JETP {\bf{19}}, 960 (1964).

\end{thebibliography}
\end{document}